\documentclass[twocolumn, pra, superscriptaddress]{revtex4}
\usepackage{amsfonts}
\usepackage{amssymb}
\usepackage{amsmath}
\usepackage{epsfig}
\usepackage{color}
\usepackage{graphics, graphicx}
\usepackage{bbold}
\usepackage{psfrag}
\usepackage{mathcomp}
\usepackage{subfigure}
\usepackage{verbatim}
\usepackage{color}
\usepackage[colorlinks,citecolor=blue]{hyperref}

\begin{document}

\title{A new method for driven-dissipative problems: Keldysh-Heisenberg
equations}
\author{Yuanwei Zhang}
\thanks{zywznl@163.com}
\affiliation{College of Physics and Electronic Engineering, Sichuan Normal University,
Chengdu 610101, China}
\author{Gang Chen}
\thanks{chengang971@163.com}
\affiliation{State Key Laboratory of Quantum Optics and Quantum Optics Devices, Institute
of Laser Spectroscopy, Shanxi University, Taiyuan, Shanxi 030006, China}
\affiliation{Collaborative Innovation Center of Extreme Optics, Shanxi University,
Taiyuan, Shanxi 030006, China}
\affiliation{Collaborative Innovation Center of Light Manipulations and Applications,
Shandong Normal University, Jinan 250358, China}

\begin{abstract}
Driven-dissipative systems have recently attracted great attention due to
the existence of novel physical phenomena with no analog in the equilibrium
case. The Keldysh path-integral theory is a powerful tool to investigate
these systems. However, it has still been challenge to study strong
nonlinear effects implemented by recent experiments, since in this case the
photon number is few and quantum fluctuations play a crucial role in
dynamics of system. Here we develop a new approach for deriving exact steady
states of driven-dissipative systems by introducing the Keldysh partition
function in the Fock-state basis and then mapping the standard saddle-point
equations into Keldysh-Heisenberg equations. We take the strong Kerr
nonlinear resonators with/without the nonlinear driving as two examples to
illustrate our method. It is found that in the absence of the nonlinear
driving, the exact steady state obtained does not exhibit bistability and
agree well with the complex \textit{P}-representation solution. While in the
presence of the nonlinear driving, the multiphoton resonance effects are
revealed and are consistent with the qualitative analysis. Our method
provides an intuitive way to explore a variety of driven-dissipative systems
especially with strong correlations.
\end{abstract}

\maketitle

\section{INTRODUCTION}

In recent years, the driven-dissipative systems have got a lot of attentions
both theoretically and experimentally. In these systems, the nonlinear
interactions can be significantly enhanced by controlling both the driving
and dissipation processes. For example, strong optical nonlinearities at the
single-photon level have already been observed in cavity quantum
electrodynamics (QED) \cite{KM05,AR13}, Rydberg atomic systems \cite%
{HG14,HB17,SH20}, optomechanical systems \cite{MA14}, and superconducting
circuit QED systems \cite{PM00,AR12,ZL15,ST18,AG19,RL20}. These advances in
experimental methods have greatly promoted the development of quantum
metrology, quantum information and quantum optical devices \cite{VG11,AB20}.
On the other hand, they also provide good platforms for studying novel
nonequilibrium physical phenomena, such as the dynamical critical phenomena
\cite{LM13,LM14,JT20}, time crystals \cite{FM19}, driven-dissipative strong
correlations \cite{TT17,RM19}. In this context, how to understand the
nonlinear effects in nonequilibrium phenomena has become an important topic.

The Keldysh functional integral formalism in the coherent-state basis is a
general approach to study nonequilibrium physics \cite{Keldysh}. This
technique provides a well-developed toolbox of perturbation techniques to
study the nonlinear effects \cite{AB10,AK11,LM16}. For example, in some
systems such as the polariton condensates \cite{LM13,LM14} and atomic
ensembles in cavity \cite{EG13,MB13,DN15,EG16,YS20,PK18}, the
single-particle actions are quadratic and the diagrammatic perturbation
theory, based on the Wick's theorem, can be performed. However, for
coherently driven systems such as optomechanical systems \cite%
{MA13,MA15,MA16,My}, the single-particle actions are no longer quadratic and
the Wick's theorem cannot be applied directly. Fortunately, when the
coherent driving is strong and the nonlinear interaction is weak, the mean
photon number is large and the standard saddle-point approximation can be
well introduced. In this approach, the mean value of operators are mainly
determined by the classical path, which satisfies the saddle-point
equations, and quantum fluctuations are treated as perturbation \cite%
{AB10,AK11,LM16}. However, recent researches have focused on the strongly
nonlinear effects at the level of individual photons, which are benefit for
processing quantum information \cite{AB20}. Experimentally, these require
the systems are weakly driven and the nonlinear interactions are strong. As
a result, the mean photon number is few and quantum fluctuations play a
crucial role in dynamics of system. This indicates that the standard
saddle-point approximation is not reasonable.

To solve this crucial problem, we develop the Keldysh path-integral theory
in the Fock-state basis, from which the standard saddle-point equations are
mapped into quantum Hamiltonian equations named as Keldysh-Heisenberg
equations. As a result, the exact steady states induced by the quantum
fluctuation effect can be well derived. We take the strong Kerr nonlinear
resonators with/without the nonlinear driving as two examples to illustrate
our method. It is found that in the absence of the nonlinear driving, the
exact steady state obtained does not exhibit bistability and agree well with
the complex \textit{P}-representation solutions. While in the presence of
the nonlinear driving, the multiphoton resonance effects are revealed and
are consistent with the qualitative analysis. Our method offers an effective
way to explore a variety of driven-dissipative systems especially with
strongly-correlated photons, based on the powerful toolbox of quantum field
theory.\newline

\begin{figure}[tb]
\includegraphics[width=7.0cm]{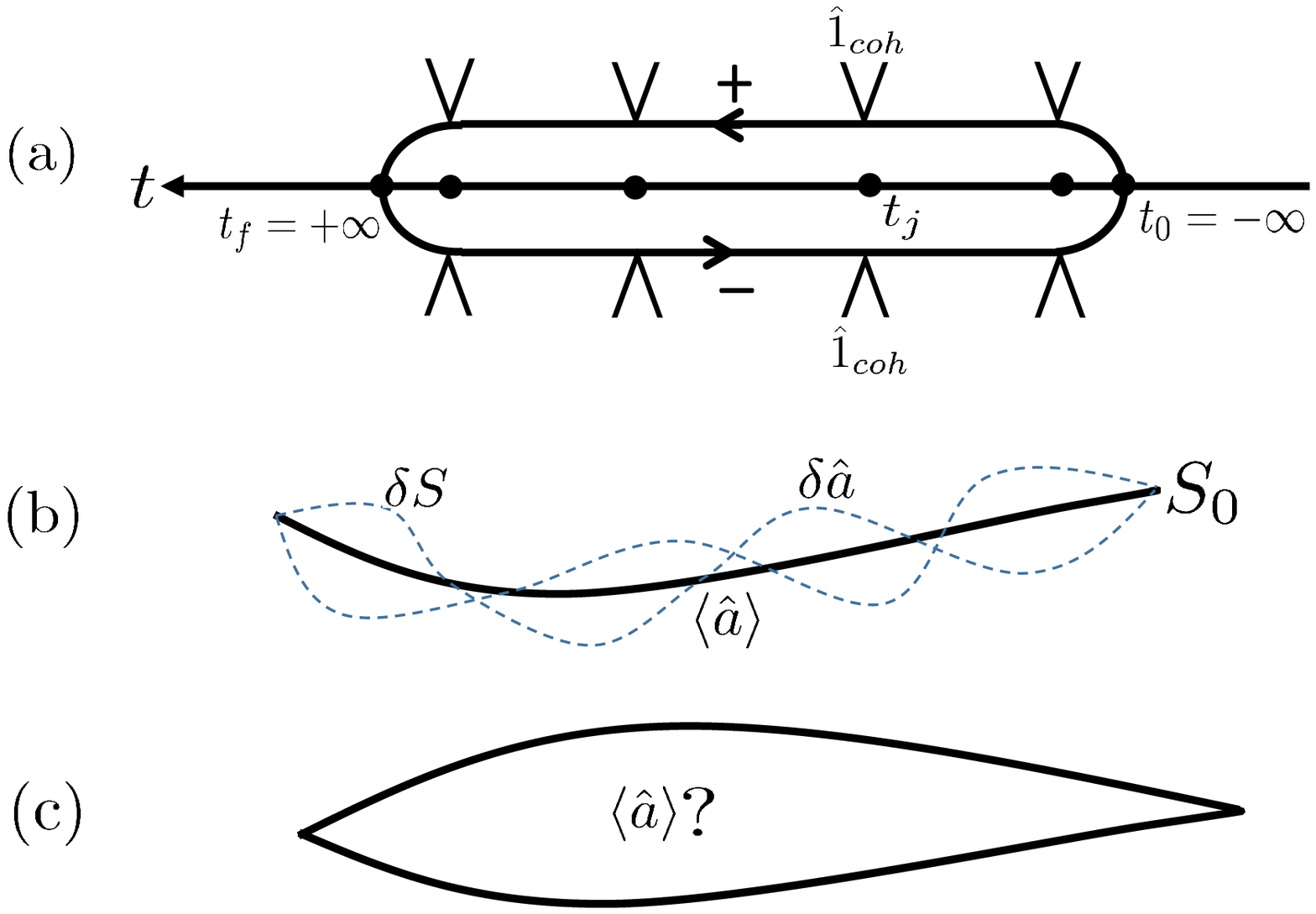}\newline
\caption{(a) The Keldysh closed time contour in the coherent-state basis ($%
\hat{1}_{coh}$). (b) Schematic diagram of the classical path (black solid
line) and its quantum fluctuations (gray dashed lines). The classical path
satisfies the saddle-point equations and has the action $S_{0}$, while the
quantum fluctuations have the action $\protect\delta S$. In the context of
quantum optics, the operator $\hat{a}$ can be split into $\hat{a}\rightarrow
\langle \hat{a}\rangle +\protect\delta \hat{a}$, where $\langle \hat{a}%
\rangle $ describes the classical path, and $\protect\delta \hat{a}$ governs
the quantum fluctuation effect. (c) When the coherent driving is weak and
the nonlinear interaction is strong, the saddle-point equations may have two
solutions. }
\label{fig1}
\end{figure}

\section{Standard saddle-point approximation}

We begin to consider a dissipative Kerr nonlinear resonator with a coherent
driving term, in which the Hamiltonian is written as ($\hbar =1$ hereafter)%
\begin{equation}
\hat{H}=\Delta _{c}\hat{a}^{\dag }\hat{a}+\chi \hat{a}^{\dag 2}\hat{a}%
^{2}+i\Omega \left( \hat{a}^{\dag }-\hat{a}\right) ,  \label{H}
\end{equation}%
where $\hat{a}$ ($\hat{a}^{\dag }$) is the annihilation (creation) operator
of the resonator, $\Delta _{c}=\omega _{c}-\omega _{p}$ is the detuning
between the resonator and driving fields, $\chi $ is the Kerr nonlinearity,
and $\Omega $ is the driving amplitude. The dynamics of such system is
described by the Lindblad master equation \cite{PD80,Walls}%
\begin{equation}
\frac{\text{d}}{\text{d}t}\hat{\rho}\left( t\right) =\mathcal{L}\hat{\rho}%
\left( t\right) =-i\left[ \hat{H},\hat{\rho}\left( t\right) \right] +\gamma
\mathcal{D}[\hat{a}]\hat{\rho}\left( t\right) .  \label{M}
\end{equation}%
where $\hat{\rho}\left( t\right) $ is the density matrix, $\mathcal{L}$ is
the Liouville superoperator, $\gamma $\ is the one-photon decay rate, and $%
\mathcal{D}[\hat{o}]\hat{\rho}\left( t\right) =\hat{o}\hat{\rho}\left(
t\right) \hat{o}^{\dag }-[\hat{o}^{\dag }\hat{o}\hat{\rho}\left( t\right) +%
\hat{\rho}\left( t\right) \hat{o}^{\dag }\hat{o}]/2$ is the standard
dissipator in the Lindblad form. Note that the dissipative evolution
corresponds to coupling the system to a zero-temperature bath \cite%
{PD80,Walls}. This Lindblad master equation can be investigated by the
Keldysh nonequilibrium quantum field theory \cite{AB10,AK11,LM16}, in which
the evolution takes place along the closed time contour.

We suppose $\left\vert \alpha \right\rangle $ as a coherent state, which is
the eigenstate of the annihilation operator $\hat{a}$ with the complex
eigenvalue $a$ (i.e., $\hat{a}\left\vert \alpha \right\rangle =a\left\vert
\alpha \right\rangle $). Note that the Keldysh close contour can be divided
into a sequence of infinitesimal time steps, as shown in Fig.~\ref{fig1}(a).
Then the completeness relation in terms of the coherent state, $\hat{1}%
_{coh}=\iint ($d$a^{\ast }$d$a/\pi )e^{-\left\vert a\right\vert
^{2}}\left\vert \alpha \right\rangle \left\langle \alpha \right\vert $, is
inserted in between consecutive time steps \cite{LM16}. In this
coherent-state basis, the partition function, which is corresponding to the
Lindblad master equation (\ref{M}), is given by%
\begin{equation}
Z=\int \mathfrak{D}\left[ a_{+},a_{-}\right] \exp \left( iS\right) ,
\end{equation}%
where $+$ and $-$ denote the forward and backward branches and the action
\begin{eqnarray}
S\!\!\! &=&\!\!\!\int\nolimits_{-\infty }^{+\infty }\text{d}t\left\{
a_{+}^{\ast }\left( i\partial _{t}-\Delta _{c}\right) a_{+}-\chi a_{+}^{\ast
2}a_{+}^{2}-i\Omega \left( a_{+}^{\ast }-a_{+}\right) \right.  \notag \\
&&\left. -a_{-}^{\ast }\left( i\partial _{t}-\Delta _{c}\right) a_{-}+\chi
a_{-}^{\ast 2}a_{-}^{2}+i\Omega \left( a_{-}^{\ast }-a_{-}\right) \right.
\notag \\
&&-i\gamma a_{+}a_{-}^{\ast }+i\frac{\gamma }{2}\left( a_{+}^{\ast
}a_{+}+a_{-}^{\ast }a_{-}\right) \}.  \label{S}
\end{eqnarray}%
It is more convenient to discuss Eq.~(\ref{S}) in the Keldysh basis,
\begin{equation}
a_{cl}=\frac{1}{\sqrt{2}}\left( a_{+}+a_{-}\right) ,\left. {}\right. \left.
{}\right. a_{q}=\frac{1}{\sqrt{2}}\left( a_{+}-a_{-}\right) ,  \label{K1}
\end{equation}%
where $a_{cl}$ and $a_{q}$ are the classical and quantum fields \cite%
{AB10,AK11,LM16}. After a straightforward calculation, the action is
rewritten as
\begin{eqnarray}
S &=&\int\nolimits_{-\infty }^{+\infty }\text{d}t\left\{ a_{cl}^{\ast
}\left( i\partial _{t}-\Delta _{c}\right) a_{q}+a_{q}^{\ast }\left(
i\partial _{t}-\Delta _{c}\right) a_{cl}\right.  \notag \\
&&\!\!\!\!\!\!\!\!\!-i\frac{\gamma }{2}\left( a_{cl}^{\ast
}a_{q}-a_{cl}a_{q}^{\ast }\right) +i\gamma a_{q}^{\ast }a_{q}-i\sqrt{2}%
\Omega \left( a_{q}^{\ast }-a_{q}\right)  \notag \\
&&\!\!\!\!\!\!\!\!\!-\chi (a_{cl}^{\ast 2}a_{cl}a_{q}+a_{cl}a_{q}^{\ast
2}a_{q}+a_{cl}^{\ast }a_{cl}^{2}a_{q}^{\ast }+a_{cl}^{\ast }a_{q}^{\ast
}a_{q}^{2})\}.  \label{S0}
\end{eqnarray}%
Note that in the presence of the coherent driving ($\Omega \neq 0$), the
first two lines of Eq.~(\ref{S0}) are not quadratic. Therefore we cannot
directly apply the diagrammatic perturbation theory, which is based on the
Wick's theorem, to calculate the nonlinear term. Fortunately, when the
coherent driving is strong and the nonlinear interaction is weak, the mean
photon number circulating inside the resonator is large and the light field
behaves as a semi-classical field \cite{Walls}. In such a case, the
saddle-point approximation can be well used to investigate the dynamics of
system \cite{AB10,AK11,LM16}. As show in Fig.~\ref{fig1}(b), the mean value
of operators are mainly determined by the classical path and the quantum
fluctuations are treated as perturbation. The classical path is determined
by the principle of least action:%
\begin{equation}
\frac{\delta S}{\delta a_{cl}^{\ast }}=0,\left. \left. {}\right. \right.
\frac{\delta S}{\delta a_{q}^{\ast }}=0,  \label{SE}
\end{equation}%
which lead to two saddle-point equations%
\begin{equation}
i\partial _{t}a_{q}=\frac{1}{2}\left( 2\Delta _{c}+i\gamma \right)
a_{q}+\chi (2a_{cl}^{\ast }a_{cl}a_{q}+a_{cl}^{2}a_{q}^{\ast }+a_{q}^{\ast
}a_{q}^{2}),  \label{SE1}
\end{equation}%
\begin{eqnarray}
i\partial _{t}a_{cl} &=&i\sqrt{2}\Omega -i\gamma a_{q}+\frac{1}{2}\left(
2\Delta _{c}-i\gamma \right) a_{cl}  \notag \\
&&+\chi (2a_{cl}a_{q}^{\ast }a_{q}+a_{cl}^{\ast }a_{cl}^{2}+a_{cl}^{\ast
}a_{q}^{2}).  \label{SE2}
\end{eqnarray}

Equation~(\ref{SE1}) is always solved by%
\begin{equation}
a_{q}=a_{q}^{\ast }=0.  \label{aq}
\end{equation}%
By substituting $a_{q}=a_{q}^{\ast }=0$ into Eq.~(\ref{S0}), we find that
the action $S=0$ in the steady-state case. Indeed, in this case, $%
a_{+}=a_{-} $ and the action on the forward part of the contour is canceled
by that on the backward part \cite{AK11}. In addition, we also obtain $%
a_{cl}=\sqrt{2}a_{0}$, where $a_{0}=a_{+}=a_{-}$ is the steady-state mean
value of $\hat{a}$, i.e., $a_{0}=\left\langle \hat{a}\right\rangle $ \cite%
{EG13}. By substituting $a_{q}=0$ and $a_{cl}=\sqrt{2}a_{0}$ into Eq.~(\ref%
{SE2}) and making $i\partial _{t}a_{cl}=0$, we obtain $a_{0}=-2i\Omega
/(2\Delta _{c}-i\gamma +4\chi \left\vert a_{0}\right\vert ^{2})$, from which
the mean photon number
\begin{equation}
\left\langle \hat{a}^{\dag }\hat{a}\right\rangle =\left\vert
a_{0}\right\vert ^{2}=\frac{4\Omega ^{2}}{4\left( \Delta _{c}+2\chi
\left\vert a_{0}\right\vert ^{2}\right) ^{2}+\gamma ^{2}}.  \label{n1}
\end{equation}

\begin{figure}[b]
\includegraphics[width=8.0cm]{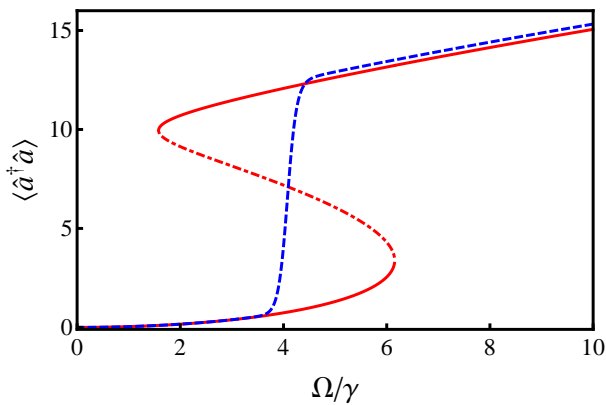}\newline
\caption{The steady-state mean photon number $\left\langle \hat{a}^{\dag }%
\hat{a}\right\rangle $ as a function of the coherent driving amplitude $%
\Omega /\protect\gamma $, when $\Delta _{c}/\protect\gamma =5$ and $\protect%
\chi /\protect\gamma =-0.25$. The red solid lines are the stable solutions
of Eq.~(\protect\ref{n1}), while the red dash-dotted line is its unstable
solution. These mean-field solutions reflect the optical bistability
phenomenon. The blue dashed line is the exact steady-state solution from
Eq.~(\protect\ref{R1}), which includes quantum fluctuations. }
\label{fig2}
\end{figure}

This solution is identical to the mean-field solution of the steady state
\cite{PD80,Walls}. In fact, the saddle-point approximation is equivalent to
the mean-field approach, named the linearization approximation, in quantum
optics \cite{Walls}. In the spirit of the linearization approximation, the
operator $\hat{a}$ can be split into an average amplitude and a fluctuation
term, i.e., $\hat{a}\rightarrow \langle \hat{a}\rangle +\delta \hat{a}$,
where $\langle \hat{a}\rangle $ is determined by the mean-field equation.
The correspondence between these two methods is shown in Fig.~\ref{fig1}(b).
As pointed out in Ref.~\cite{PD80}, when the sign of $\Delta _{c}$ is
opposite to that of $\chi $, Eq.~(\ref{n1}) may have two stable solutions;
see the red lines of Fig.~\ref{fig2}. In other words, the action of system
has two classical paths; see Fig.~\ref{fig1}(c). As a result, the
perturbation calculation around the classical path maybe not reasonable.
This phenomenon, named as optical bistability, signals the failure of both
the linearization approximation and the saddle-point approximation. On the
contrary, Drummond and Walls derived a complex \textit{P}-representation
solution for the steady state \cite{PD80}. In that method, they considered
the quantum fluctuation effect and found that the exact steady-state
solution does not exhibit bistability.\newline

\begin{figure}[tb]
\includegraphics[width=6.0cm]{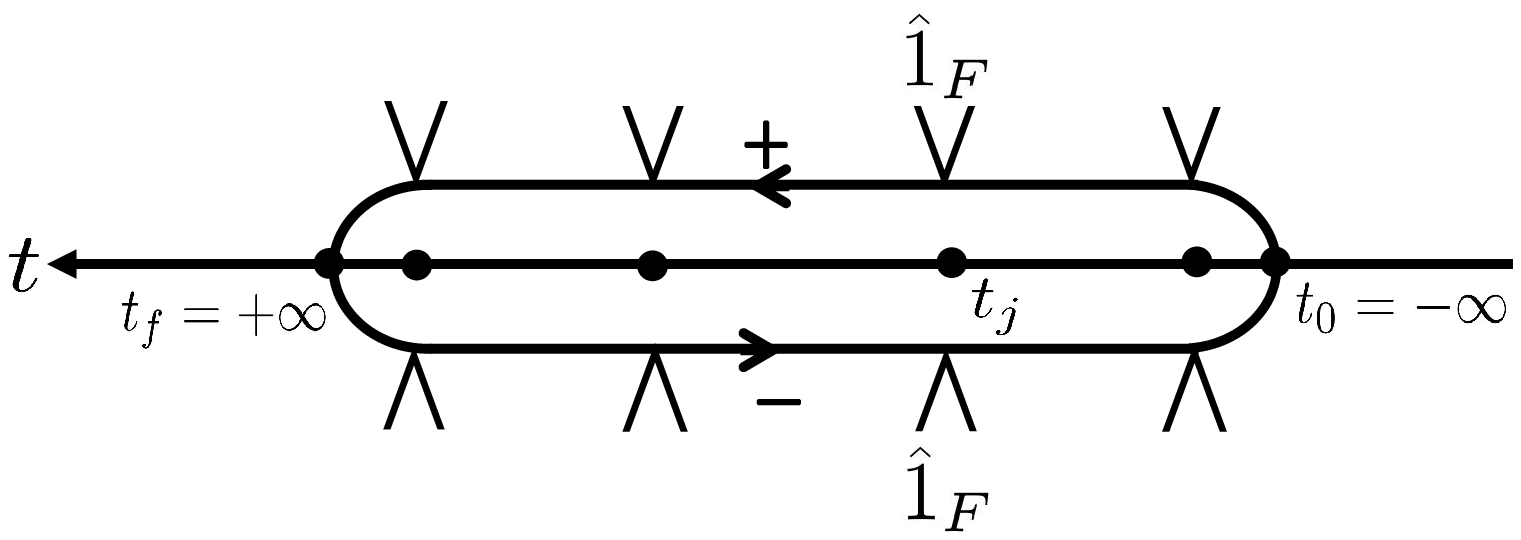}\newline
\caption{The Keldysh closed time contour in the Fock-state basis ($\hat{1}%
_{F}=\sum\nolimits_{n}\left\vert n\right\rangle \left\langle n\right\vert $%
). }
\label{fig3}
\end{figure}

\section{Keldysh-Heisenberg equations}

We note that the standard saddle-point equations are based on the
coherent-state basis. The coherent state is the closest quantum mechanical
state to a classical description of the field. It is a suitable
representation for optical fields when the total photon number is large and
the quantum fluctuations are weak \cite{Walls}. Obviously, this condition is
not satisfied in the bistable region. As shown in Fig.~\ref{fig2}, in that
region the coherent driving is weak and the Kerr nolinearity is the same
order as the other parameters. Therefore the mean photon number is not so
large and quantum fluctuations induced by the Kerr nonlinearity can not be
ignored. To overcome this shortcoming, we introduce the Fock state, which is
the eigenstate of the photon number operator. In the Fock-state basis, we
develop a new method called Keldysh-Heisenberg equations that governs the
quantum fluctuation effect.

In the Fock-state basis, the completeness relation inserted in between
consecutive time steps of the Keldysh close time contour becomes $\hat{1}%
_{F}=\sum\nolimits_{n}\left\vert n\right\rangle \left\langle n\right\vert $;
see Fig.~\ref{fig3}. In this case, the Keldysh partition function for
stationary states reads (see Appendix A for details)
\begin{equation}
Z=\text{Tr}\left[ \exp (i\hat{S})\right] ,  \label{Zq}
\end{equation}%
where Tr denotes the trace operation which connects the two time branches,
giving rise to the closed Keldysh contour \cite{LM16}. $\hat{S}$ is the
quantum action (like a time evolution operator) that%
\begin{equation}
\hat{S}=-\int\nolimits_{-\infty }^{+\infty }\mathcal{\hat{H}}\text{d}t.
\label{Sq}
\end{equation}%
In Eq.~(\ref{Sq}), $\mathcal{\hat{H}}$ is a generalized Hamiltonian
operator. As shown in Appendix A, $\mathcal{\hat{H}}$ consists of operators
acting on different branches of the Keldysh closed time contour. For the
driven-dissipative Kerr nonlinear resonator described in Eq.~(\ref{M}), the
generalized Hamiltonian operator has the form
\begin{eqnarray}
\mathcal{\hat{H}} &=&\Delta _{c}\hat{a}_{+}^{\dag }\hat{a}_{+}+\chi \hat{a}%
_{+}^{\dag 2}\hat{a}_{+}^{2}+i\Omega \left( \hat{a}_{+}^{\dag }-\hat{a}%
_{+}\right)  \notag \\
&&-\Delta _{c}\hat{a}_{-}^{\dag }\hat{a}_{-}-\chi \hat{a}_{-}^{\dag 2}\hat{a}%
_{-}^{2}-i\Omega \left( \hat{a}_{-}^{\dag }-\hat{a}_{-}\right)  \notag \\
&&+i\gamma \hat{a}_{+}\hat{a}_{-}^{\dag }-i\frac{\gamma }{2}(\hat{a}%
_{+}^{\dag }\hat{a}_{+}+\hat{a}_{-}^{\dag }\hat{a}_{-}),
\end{eqnarray}%
where $\hat{a}_{\pm }$ ($\hat{a}_{\pm }^{\dag }$) are the annihilation
(creation) operators and the subscript $+$ ($-$) means that the operator
only acts on the forward (backward) time branch. Note that these operators
obey the commutation relations: $[\hat{a}_{+},\hat{a}_{+}^{\dag }]=[\hat{a}%
_{-},\hat{a}_{-}^{\dag }]=1$ and $[\hat{a}_{+},\hat{a}_{-}]=0$. Comparing
Eq.~(\ref{Sq}) with Eq.~(\ref{S}), it can be found that we only need to
replace the complex variable $a_{+}$\ ($a_{-}$) by the corresponding
operator $\hat{a}_{+}$\ ($\hat{a}_{-}$) and omit the derivative with respect
to time. We further define $\hat{a}_{cl}=(\hat{a}_{+}+\hat{a}_{-})/\sqrt{2}$
and $\hat{a}_{q}=(\hat{a}_{+}-\hat{a}_{-})/\sqrt{2}$ as the annihilation
operators of the classical and quantum fields, respectively. Immediately,
these operators obey the commutation relations: $[\hat{a}_{cl},\hat{a}%
_{cl}^{\dag }]=[\hat{a}_{q},\hat{a}_{q}^{\dag }]=1$\ and $[\hat{a}_{q},\hat{a%
}_{cl}]=0$. And the quantum action in Eq.~(\ref{Sq}) is changed to $\hat{S}%
=-\int\nolimits_{-\infty }^{+\infty }\hat{H}$d$t=-\int\nolimits_{-\infty
}^{+\infty }\left( \mathcal{\hat{H}}_{\uparrow }+\mathcal{\hat{H}}%
_{\downarrow }\right) $d$t$, where%
\begin{eqnarray}
\mathcal{\hat{H}}_{\uparrow } &=&i\sqrt{2}\Omega \hat{a}_{q}^{\dag }+\frac{1%
}{2}\left( 2\Delta _{c}-i\gamma \right) \hat{a}_{q}^{\dag }\hat{a}_{cl}
\notag \\
&&+\chi \left( \hat{a}_{cl}^{\dag }\hat{a}_{cl}+\hat{a}_{q}^{\dag }\hat{a}%
_{q}-1\right) \hat{a}_{q}^{\dag }\hat{a}_{cl},  \label{Hc}
\end{eqnarray}%
\begin{eqnarray}
\mathcal{\hat{H}}_{\downarrow } &=&-i\sqrt{2}\Omega \hat{a}_{q}-i\gamma \hat{%
a}_{q}^{\dag }\hat{a}_{q}+\frac{1}{2}\left( 2\Delta _{c}+i\gamma \right)
\hat{a}_{cl}^{\dag }\hat{a}_{q}  \notag \\
&&+\chi \left( \hat{a}_{cl}^{\dag }\hat{a}_{cl}+\hat{a}_{q}^{\dag }\hat{a}%
_{q}-1\right) \hat{a}_{cl}^{\dag }\hat{a}_{q}.  \label{Hq}
\end{eqnarray}%
Interestingly, the saddle-point equations (\ref{SE1}) and (\ref{SE2})
correspond to the following quantum Hamiltonian equations:%
\begin{equation}
i\frac{\text{d}}{\text{d}t}\hat{a}_{q}=\left[ \hat{a}_{cl},\mathcal{\hat{H}}%
\right] ,\left. {}\right. \left. {}\right. i\frac{\text{d}}{\text{d}t}\hat{a}%
_{cl}=\left[ \hat{a}_{q},\mathcal{\hat{H}}\right] .  \label{QS}
\end{equation}%
Since the equations in Eq.~(\ref{QS}) are formally similar to the Heisenberg
equations for an equilibrium system, we can call them Keldysh-Heisenberg
equations. Compared with the standard saddle-point equations, these operator
equations can completely capture the information induced by quantum
fluctuations. Therefore, we use them to obtain the exact steady-state
solution.

We first rewrite Eq.~(\ref{QS}) as a generalized Schr\"{o}dinger equation%
\begin{equation}
i\partial _{t}\left\vert \Psi _{0}\right\rangle =\mathcal{\hat{H}}\left\vert
\Psi _{0}\right\rangle ,  \label{Sch}
\end{equation}%
where $\left\vert \Psi _{0}\right\rangle $ is the steady-state wave function
and can be formally expressed as $\left\vert \Psi _{0}\right\rangle
=\left\vert \psi \right\rangle _{q}\otimes \left\vert \psi \right\rangle
_{cl}$. Note that in the steady state, the action on the forward part of the
contour is canceled by that on the backward part. This means $\hat{a}%
_{+}\left\vert \Psi _{0}\right\rangle =$\ $\hat{a}_{-}\left\vert \Psi
_{0}\right\rangle $,\ and the steady-state wave function must be the vacuum
state of $\hat{a}_{q}$, i.e., $\hat{a}_{q}\left\vert \Psi _{0}\right\rangle
=0$. This condition is corresponding to the mean-field saddle-point solution
$a_{q}=0$\ in Eq.~(\ref{aq}). As a result, the steady-state wave function is
assumed as%
\begin{equation}
\left\vert \Psi _{0}\right\rangle =\left\vert 0\right\rangle
_{q}\sum\nolimits_{m=0}^{+\infty }\beta _{m}\left\vert m\right\rangle _{cl},
\label{SW}
\end{equation}%
where $\left\vert m\right\rangle _{cl}$ is the Fock state in the occupation
number basis and $\beta _{m}$ is the expansion coefficient. Interestingly,
using Eqs.~(\ref{Hc}), (\ref{Hq}) and (\ref{SW}), we verify $\mathcal{\hat{H}%
}\left\vert \Psi _{0}\right\rangle =0$. In other words, the steady state is
the ground state of $\mathcal{\hat{H}}$. At the same time, it is straightly
to see $\hat{S}\left\vert \Psi _{0}\right\rangle =-\int\nolimits_{-\infty
}^{+\infty }\mathcal{\hat{H}}dt\left\vert \Psi _{0}\right\rangle =0$, which
is corresponding to the above discussion in section II that $S=0$ in
steady-state case.

Finally, using $\mathcal{\hat{H}}\left\vert \Psi _{0}\right\rangle =0$, we
get a recursion relation for the expansion coefficient as
\begin{equation}
\beta _{m}=\sqrt{\frac{2}{m}}\frac{\epsilon }{x+m-1}\beta _{m-1},  \label{r}
\end{equation}%
with $\epsilon =\Omega /\left( i\chi \right) $ and $x=(2\Delta _{c}-i\gamma
)/\left( 2\chi \right) $. Based on this recursion relation, the steady-state
wave function
\begin{equation}
\left\vert \Psi _{0}\right\rangle =\frac{1}{\sqrt{N}}\left\vert
0\right\rangle _{q}\sum\limits_{m=0}^{+\infty }\frac{(\sqrt{2}\epsilon )^{m}%
}{\sqrt{m!}}\frac{\Gamma (x)}{\Gamma (x+m)}\left\vert m\right\rangle _{cl},
\label{WF1}
\end{equation}%
where $\Gamma (x)$ is the gamma special function and $N=\left.
_{0}F_{2}(x^{\ast },x;2\left\vert \epsilon \right\vert ^{2})\right. $ is the
normalization constant with $\left. _{0}F_{2}(x^{\ast },x;2\left\vert
\epsilon \right\vert ^{2})\right. =\sum\nolimits_{m=0}^{+\infty }\frac{%
\Gamma (x^{\ast })\Gamma (x)}{\Gamma (x^{\ast }+m)\Gamma (x+m)}\frac{%
(2\left\vert \epsilon \right\vert ^{2})^{m}}{m!}$ being the generalized
hypergeometric function. According to the relation $\hat{a}_{cl}=(\hat{a}%
_{+}+\hat{a}_{-})/\sqrt{2}$, we obtain the steady-state correlation function
\begin{eqnarray}
\left\langle \hat{a}^{\dag l}\hat{a}^{k}\right\rangle &=&\left\langle \hat{a}%
_{+}^{\dag l}\hat{a}_{+}^{k}\right\rangle =\frac{1}{\sqrt{2^{l+k}}}%
\left\langle \Psi _{0}\right\vert \hat{a}_{cl}^{\dag l}\hat{a}%
_{cl}^{k}\left\vert \Psi _{0}\right\rangle  \label{R1} \\
&=&\frac{(\epsilon ^{\ast })^{l}\epsilon ^{k}\Gamma (x^{\ast })\Gamma (x)}{%
\Gamma (x^{\ast }+l)\Gamma (x+k)}\frac{_{0}F_{2}(x^{\ast }+l,x+k;2\left\vert
\epsilon \right\vert ^{2})}{_{0}F_{2}(x^{\ast },x;2\left\vert \epsilon
\right\vert ^{2})},  \notag
\end{eqnarray}%
which is equivalent to the complex \textit{P-}representation solution in
Ref.~\cite{PD80}. In Fig.~\ref{fig2}, we plot the steady-state mean photon
number $\left\langle \hat{a}^{\dag }\hat{a}\right\rangle $ as a function of
the coherent driving amplitude $\Omega $. Obviously, the exact steady state
does not exhibit bistability; see the blue dashed line.\newline

\section{Nonlinear driving case}

In this section, we extend our method to the two-photon nonlinear driving
case implemented recently in superconducting quantum circuits \cite{VV19}.
The effective Hamiltonian reads
\begin{equation}
\hat{H}\!=\!\Delta _{c}\hat{a}^{\dag }\hat{a}+\chi \hat{a}^{\dag 2}\hat{a}%
^{2}+i\Omega \left( \hat{a}^{\dag }\!-\!\hat{a}\right) +\frac{1}{2}\!\left(
\Lambda \hat{a}^{\dag 2}\!+\!\Lambda ^{\ast }\hat{a}^{2}\right) ,  \label{HN}
\end{equation}%
where $\Lambda $ is the complex amplitude of the two-photon driving term.
The Lindblad master equation becomes%
\begin{equation}
\frac{\text{d}}{\text{d}t}\hat{\rho}\left( t\right) =-i\left[ \hat{H},\hat{%
\rho}\left( t\right) \right] +\gamma \mathcal{D}\left[ \hat{a}\right] \hat{%
\rho}\left( t\right) +\kappa \mathcal{D}\left[ \hat{a}^{2}\right] \hat{\rho}%
\left( t\right) ,
\end{equation}%
where $\kappa $ is the two-photon loss rate.

\begin{figure}[b]
\includegraphics[width=7.5cm]{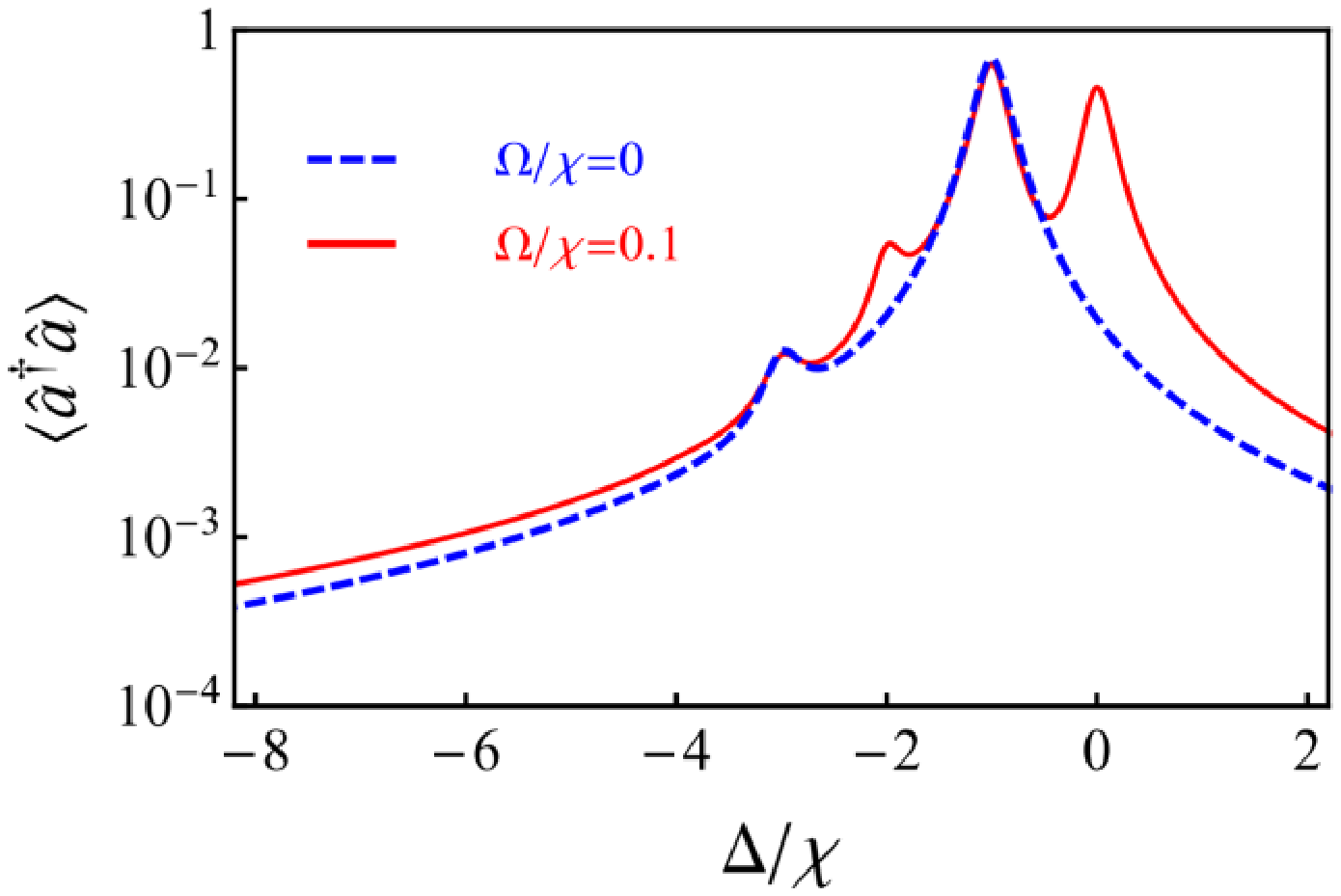}\newline
\caption{The steady-state mean photon number $\left\langle \hat{a}^{\dag }%
\hat{a}\right\rangle $ as a function of the detuning $\Delta _{c}/\protect%
\chi $, when $\Omega /\protect\chi =0$ (blue dashed line) and $\Omega /%
\protect\chi =0.1$ (red solid line). The other parameters are chosen as $%
\protect\gamma /\protect\chi =0.1$, $\protect\kappa /\protect\chi =0.1$, and
$\Lambda /\protect\chi =0.2$.}
\label{fig4}
\end{figure}

In the presence of the two-photon driving and loss terms, we rewrite $%
\mathcal{\hat{H}=\hat{H}}_{\uparrow }+\mathcal{\hat{H}}_{\downarrow }$ as
\begin{eqnarray}
\mathcal{\hat{H}}_{\uparrow } &=&\frac{1}{2}\left( 2\Delta _{c}-i\gamma
\right) \hat{a}_{q}^{\dag }\hat{a}_{cl}+\chi \left( \hat{a}_{cl}^{\dag }\hat{%
a}_{cl}+\hat{a}_{q}^{\dag }\hat{a}_{q}-1\right) \hat{a}_{q}^{\dag }\hat{a}%
_{cl}  \notag \\
&&+i\sqrt{2}\Omega \hat{a}_{q}^{\dag }-i\frac{\kappa }{2}\left( \hat{a}%
_{cl}^{\dag }\hat{a}_{cl}-\hat{a}_{q}^{\dag }\hat{a}_{q}+1\right) \hat{a}%
_{q}^{\dag }\hat{a}_{cl}  \notag \\
&&+\Lambda \hat{a}_{q}^{\dag }\hat{a}_{cl}^{\dag },  \label{Hc1}
\end{eqnarray}%
\begin{eqnarray}
\mathcal{\hat{H}}_{\downarrow } &=&\frac{1}{2}\left( 2\Delta _{c}+i\gamma
\right) \hat{a}_{cl}^{\dag }\hat{a}_{q}+\chi \left( \hat{a}_{cl}^{\dag }\hat{%
a}_{cl}+\hat{a}_{q}^{\dag }\hat{a}_{q}-1\right) \hat{a}_{cl}^{\dag }\hat{a}%
_{q}  \notag \\
&&-i\sqrt{2}\Omega \hat{a}_{q}+i\frac{\kappa }{2}\left( \hat{a}_{cl}^{\dag }%
\hat{a}_{cl}-\hat{a}_{q}^{\dag }\hat{a}_{q}+1\right) \hat{a}_{cl}^{\dag }%
\hat{a}_{q}  \notag \\
&&-(i\gamma +2i\kappa \hat{a}_{cl}^{\dag }\hat{a}_{cl})\hat{a}_{q}^{\dag }%
\hat{a}_{q}+\Lambda ^{\ast }\hat{a}_{cl}\hat{a}_{q}.  \label{Hq1}
\end{eqnarray}%
As shown in section II, we define the steady-state wave function $\left\vert
\Psi _{0}\right\rangle =\left\vert 0\right\rangle
_{q}\sum\nolimits_{m=0}^{\infty }\beta _{m}\left\vert m\right\rangle _{cl}$.
The condition $\mathcal{\hat{H}}\left\vert \Psi _{0}\right\rangle =0$
induces a recursion relation for the expansion coefficient as%
\begin{eqnarray}
&&\left[ \left( 2\Delta _{c}-i\gamma \right) +\left( 2\chi -i\kappa \right)
\left( m-1\right) \right] \sqrt{m}\beta _{m}  \notag \\
&&\left. =\right. -i2\sqrt{2}\Omega \beta _{m-1}-2\Lambda \sqrt{m-1}\beta
_{m-2}.  \label{r1}
\end{eqnarray}%
The last term in Eq.~(\ref{r1}), which corresponds to the term $\Lambda \hat{%
a}_{q}^{\dag }\hat{a}_{cl}^{\dag }$ in $\mathcal{\hat{H}}_{\uparrow }$,
makes the recursion relation difficult to solve. To simplify the
calculation, we use a displacement transformation $\mathcal{\hat{H}}%
^{^{\prime }}=e^{\lambda \hat{a}_{cl}^{\dag }}\left( \mathcal{\hat{H}}%
_{\uparrow }+\mathcal{\hat{H}}_{\downarrow }\right) e^{-\lambda \hat{a}%
_{cl}^{\dag }}=\mathcal{\hat{H}}_{\uparrow }^{^{\prime }}+\mathcal{\hat{H}}%
_{\downarrow }^{^{\prime }}$. Under this transformation, $\hat{a}_{q}$ does
not be changed ($e^{\lambda \hat{a}_{cl}^{\dag }}\hat{a}_{q}e^{-\lambda \hat{%
a}_{cl}^{\dag }}\rightarrow \hat{a}_{q}$), but $\hat{a}_{cl}$ has a
displacement ($e^{\lambda \hat{a}_{cl}^{\dag }}\hat{a}_{cl}e^{-\lambda \hat{a%
}_{cl}^{\dag }}\rightarrow \hat{a}_{cl}-\lambda $). When choosing $\lambda =i%
\sqrt{2\Lambda /(2\chi -i\kappa )}$, the term $\Lambda \hat{a}_{q}^{\dag }%
\hat{a}_{cl}^{\dag }$ can be eliminated and the condition $\mathcal{\hat{H}}%
\left\vert \Psi _{0}\right\rangle =0$ is thus equivalent to $\mathcal{\hat{H}%
}^{^{\prime }}\left\vert \Phi _{0}\right\rangle =0$ with $\left\vert \Phi
_{0}\right\rangle =e^{\lambda \hat{a}_{cl}^{\dag }}\left\vert \Psi
_{0}\right\rangle =\left\vert 0\right\rangle _{q}\sum\nolimits_{m=0}^{\infty
}\phi _{m}\left\vert m\right\rangle _{cl}$, where $\phi _{m}$ is also the
expansion coefficient. Since $\mathcal{\hat{H}}_{\downarrow }$ is
proportional to $\hat{a}_{q}$, the equation $\mathcal{\hat{H}}^{^{\prime
}}\left\vert \Phi _{0}\right\rangle =0$ reduces to $\mathcal{\hat{H}}%
_{\uparrow }^{^{\prime }}\left\vert \Phi _{0}\right\rangle =0$, where%
\begin{eqnarray}
\mathcal{\hat{H}}_{\uparrow }^{^{\prime }} &=&\!\!\!\chi \left[ \hat{a}%
_{cl}^{\dag }\hat{a}_{cl}\hat{a}_{cl}-2\lambda \hat{a}_{cl}^{\dag }\hat{a}%
_{cl}+\left( \hat{a}_{cl}-\lambda \right) \left( \hat{a}_{q}^{\dag }\hat{a}%
_{q}-1\right) \right] \hat{a}_{q}^{\dag }  \notag \\
&&\!\!\!\!\!\!\!\!\!-i\frac{\kappa }{2}\left[ \hat{a}_{cl}^{\dag }\hat{a}%
_{cl}\hat{a}_{cl}-2\lambda \hat{a}_{cl}^{\dag }\hat{a}_{cl}-\left( \hat{a}%
_{cl}-\lambda \right) \left( \hat{a}_{q}^{\dag }\hat{a}_{q}-1\right) \right]
\hat{a}_{q}^{\dag }  \notag \\
&&\!\!\!\!\!\!\!\!\!+i\sqrt{2}\Omega \hat{a}_{q}^{\dag }+\frac{1}{2}(2\Delta
_{c}-i\gamma )\left( \hat{a}_{cl}-\lambda \right) \hat{a}_{q}^{\dag }.
\end{eqnarray}%
And the recursion relation for the expansion coefficient is given by
\begin{equation}
\phi _{m}=\frac{2\lambda }{\sqrt{m}}\frac{y+m-1}{z+m-1}\phi _{m-1},
\end{equation}%
where $y=[-i2\sqrt{2}\Omega +\lambda (2\Delta _{c}-i\gamma )]/\left[
2\lambda (2\chi -i\kappa )\right] $ and $z=(2\Delta _{c}-i\gamma )/(2\chi
-i\kappa )$. This recursion relation is solved by $\phi _{m}=\frac{\left(
2\lambda \right) ^{m}}{\sqrt{m!}}\frac{\Gamma (y+m)}{\Gamma (z+m)}$, from
which the steady-state wave function $\left\vert \Psi _{0}\right\rangle
=e^{-\lambda \hat{a}_{cl}^{\dag }}\left\vert \Phi _{0}\right\rangle $ is
obtained by (see Appendix B for details)
\begin{equation}
\left\vert \Psi _{0}\right\rangle \!\!=\!\!\frac{1}{\sqrt{N}}\left\vert
0\right\rangle _{q}\sum\limits_{m=0}^{+\infty }(-\lambda )^{m}\frac{%
_{2}F_{1}(-m,y;z;2)}{\sqrt{m!}}\left\vert m\right\rangle _{cl},  \label{WF2}
\end{equation}%
where $N=\sum\nolimits_{m=0}^{+\infty }\frac{\left\vert \lambda \right\vert
^{2m}}{m!}\left\vert _{2}F_{1}(-m,y;z;2)\right\vert ^{2}$ is the
normalization constant and $_{2}F_{1}(-m,y;z;2)=\sum\nolimits_{n=0}^{+\infty
}\frac{\left( -m\right) _{n}\left( y\right) _{n}}{\left( z\right) _{n}}\frac{%
2^{n}}{n!}$ is the generalized hypergeometric function with $\left( r\right)
_{n}=\Gamma (r+n)/\Gamma (r)$. Based on Eq.~(\ref{WF2}), the steady-state
correlation function%
\begin{equation}
\left\langle \hat{a}^{\dag l}\hat{a}^{k}\right\rangle =\frac{1}{N\sqrt{%
2^{l+k}}}\sum\limits_{m=0}^{+\infty }\frac{1}{m!}\mathcal{F}_{m+l}^{\ast }%
\mathcal{F}_{m+k}  \label{R2}
\end{equation}%
where $\mathcal{F}_{m+k}=(-\lambda )^{m+k}\left. _{2}F_{1}\left[ -(m+k),y;z;2%
\right] \right. $. It can be verified that Eq.~(\ref{R2}) is equivalent to
the solution in Ref.~\cite{NB16}.

Using Eq.~(\ref{R2}), we can study the influence of different driving
processes on the nonlinear effects. For example, we consider the multiphoton
resonances in the weak driving regime, which are easy to observe in
experiments. In this situation, the mean photon number is small and the
mean-field approach is not reasonable. We firstly make a qualitative
prediction from the Hamiltonian (\ref{HN}). When the energy of $n$ incident
photons is equivalent to the energy of $n$ photons inside the resonator,
that is $n\omega _{p}=n\omega _{c}+\chi n\left( n-1\right) $, the absorption
of $n$ pumping photons is favored. Expressed in terms of the detuning $%
\Delta _{c}=\omega _{c}-\omega _{p}$, this relation reads $\Delta _{c}/\chi
=-(n-1)$. On the other hand, the parity of $n$ depends on the driving
processes. In the absence of the one-photon driving ($\Omega =0$ and $%
\Lambda \neq 0$), even number of pumping photons are favored ($n$ is even)
and $\Delta _{c}/\chi =$ $-1,-3,-5,\cdots $, while in the presence of both
the one- and two-photon driving ($\Omega \neq 0$ and $\Lambda \neq 0$), $n$
can be any integer greater than $0$ and $\Delta _{c}/\chi
=0,-1,-2,-3,-4,\cdots $. In Fig.~\ref{fig4}, we plot the steady-state mean
photon number $\left\langle \hat{a}^{\dag }\hat{a}\right\rangle $ as a
function of the detuning $\Delta _{c}/\chi $, based on Eq.~(\ref{R2}). This
figure shows clearly that in the absence of the one-photon pumping (see the
blue dashed line), the photon resonances arise around $\Delta _{c}/\chi =$ $%
-1$ and $-3$, while in the presence of both the one- and two-photon drivings
(see the red solid line), the photon resonances arise around $\Delta
_{c}/\chi =0$, $-1$, $-2$, and $-3$. These results are consistent with the
qualitative analysis.\newline

\section{CONCLUSIONS}

In summary, we have established the Keldysh path-integral theory in the
Fock-states basis, from which the Keldysh-Heisenberg equations are
successfully introduced. In contrast to the standard saddle-point equations,
these quantum operator equations can well describe the quantum fluctuation
effect and thus present the exact steady-state solutions. We have also
considered two examples about the driven-dissipative Kerr nonlinear
resonators with/without the two-photon nonlinear driving. Our results agree
well with the qualitative analysis and those obtained by the complex \textit{%
P}-representation method \cite{PD80,NB16} and the coherent quantum-absorber
method \cite{KS12,DR20}.

Before ending up this paper, we compare our method with the complex \textit{P%
}-representation method \cite{PD80} and the coherent quantum-absorber method
\cite{KS12,DR20}, both of which have also considered the quantum fluctuation
effect. For the complex \textit{P}-representation method, an operator master
equation has been transformed to a c-number Fokker-Planck equation, and many
complicated integration operations have to be faced. While for the coherent
quantum-absorber method, a auxiliary resonator, which has the same Hilbert
space dimension as the original resonator, should be introduced. By
constructing the Hamiltonian for the auxiliary resonator appropriately, this
cascaded system has a \textquotedblleft dark\textquotedblright\ state. Then,
one can get the steady state of the original system by tracing out the
auxiliary cavity. Our developed Keldysh functional-integral method with the
Keldysh-Heisenberg equations is more physical and intuitive. Moreover, it
can be extended to deal with more complex problems such as
strongly-correlated photons \cite{RM19,AL13}, based on the powerful toolbox
of quantum field theory.\newline

\acknowledgments

This work is supported by the National Key R \& D Program of China under
Grant No.~2017YFA0304203, the National Natural Science Foundation of China
under Grant No.~11804241, and Shanxi ``1331 Project" Key Subjects
Construction.

\vbox{\vskip0cm} \appendix\setcounter{figure}{0} \renewcommand{\thefigure}{B%
\arabic{figure}}

\section{Keldysh partition function in the Fock-states basis}

In this appendix, we drive the Keldysh partition function in the Fock-states
basis in details. A general Lindblad master equation reads%
\begin{equation}
\frac{\text{d}}{\text{d}t}\hat{\rho}\left( t\right) =\mathcal{L}\hat{\rho}%
\left( t\right) =-i\left[ \hat{H},\hat{\rho}\left( t\right) \right] +\gamma
\mathcal{D}[\hat{o}]\hat{\rho}\left( t\right) .  \label{Mm}
\end{equation}%
where $\hat{H}$ is the any Hamiltonian of system. For one or two photon loss
process, $\hat{o}$ can be chosen as $\hat{a}$ or $\hat{a}^{2}$. Without loss
of generality, we set $\hat{o}=$ $\hat{a}$ hereafter. Using the master
equation (\ref{Mm}), the time evolution of the density matrix from $t_{0}$
to $t_{f}$ is formally solved by
\begin{equation}
\hat{\rho}\left( t_{f}\right) =e^{\left( t_{f}-t_{0}\right) \mathcal{L}}\hat{%
\rho}\left( t_{0}\right) =\lim_{N\rightarrow \infty }\left( 1+\delta t%
\mathcal{L}\right) ^{N}\hat{\rho}\left( t_{0}\right) ,  \label{M1}
\end{equation}%
where we have decomposed the time evolution into a sequence of infinitesimal
steps of duration $\delta t=\left( t_{f}-t_{0}\right) /N$. We focus on a
single time step, and denote the density matrix after the $j$-th step ($%
t_{j}=t_{0}+j\delta t$) by $\hat{\rho}_{j}$ $=\hat{\rho}(t_{j})$. Then we
have
\begin{equation}
\hat{\rho}_{j+1}=e^{\delta t\mathcal{L}}\hat{\rho}_{j}=\left( 1+\delta t%
\mathcal{L}\right) \hat{\rho}_{j}+O(\delta t^{2}).  \label{M2}
\end{equation}

Since the Liouville superoperator $\mathcal{L}$ acts on the density matrix
\textquotedblleft from both sides\textquotedblright . It is more convenient
to represent the density matrix in the Keldysh closed time contour. As shown
in Fig.~\ref{fig3} of the main text, this can be achieved by projecting the
density matrix into the two time branches \cite{LM16}:
\begin{equation}
\hat{\rho}_{j}\equiv \hat{P}_{+,j}\hat{\rho}_{j}\hat{P}_{-,j},  \label{M3}
\end{equation}%
where $\hat{P}_{+,j}$ ($\hat{P}_{-,j}$) is the projection operator on the
forward (backward) branch at the time $t_{j}$. Obviously, if we choose $\hat{%
P}$ as a unit operator of the coherent state, i.e., $\hat{P}=\hat{1}%
_{coh}=\iint ($d$a^{\ast }$d$a/\pi )e^{-\left\vert a\right\vert
^{2}}\left\vert \alpha \right\rangle \left\langle \alpha \right\vert $, we
can get the partition function in Sec.~II \cite{LM16}. Instead, here we
choose $\hat{P}$ as the identity in the Fock space, i.e., $\hat{P}_{\pm }=%
\hat{1}_{F}=\sum_{n}\left\vert n_{\pm }\right\rangle \left\langle n_{\pm
}\right\vert $. In this case, $\hat{\rho}_{j}$ can be written as%
\begin{eqnarray}
\hat{\rho}_{j} &\equiv &\sum\limits_{m,n}\left\vert m_{+}\right\rangle
\left\langle m_{+}\right\vert \hat{\rho}_{j}\left\vert n_{-}\right\rangle
\left\langle n_{-}\right\vert  \notag \\
&=&\sum\limits_{m,n}\left\langle m_{+}\right\vert \hat{\rho}_{j}\left\vert
n_{-}\right\rangle \left\vert m_{+}\right\rangle \left\langle
n_{-}\right\vert .  \label{M4}
\end{eqnarray}

We now consider $\hat{\rho}_{j+1}\equiv \sum\nolimits_{k,l}\left\langle
k_{+}\right\vert \hat{\rho}_{j+1}\left\vert l_{-}\right\rangle \left\vert
k_{+}\right\rangle \left\langle l_{-}\right\vert $ in terms of the
corresponding matrix element at the previous time step $t_{j}$. Inserting
Eq.~(\ref{M4}) into Eq.~(\ref{M2}), we obtain
\begin{eqnarray}
&&\!\!\!\!\!\!\!\!\!\!\!\!\left\langle k_{+}\right\vert \hat{\rho}%
_{j+1}\left\vert l_{-}\right\rangle  \notag \\
&=&\sum\limits_{m,n}\left\langle k_{+}\right\vert \left[ \left( 1+\delta t%
\mathcal{L}\right) (\left\langle m_{+}\right\vert \hat{\rho}_{j}\left\vert
n_{-}\right\rangle \left\vert m_{+}\right\rangle \left\langle
n_{-}\right\vert )\right] \left\vert l_{-}\right\rangle  \notag \\
&=&\sum\limits_{m,n}\left( \delta _{k,m}\delta _{l,n}-i\delta t\mathcal{H}%
_{k,l,m,n}\right) \left\langle m_{+}\right\vert \hat{\rho}_{j}\left\vert
n_{-}\right\rangle ,
\end{eqnarray}%
where%
\begin{eqnarray}
&&\!\!\!\!\!\!\!\!\!\!\!\!\mathcal{H}_{k,l,m,n}  \notag \\
&=&i\left\langle k_{+}\right\vert \mathcal{L}(\left\vert m_{+}\right\rangle
\left\langle n_{-}\right\vert )\left\vert l_{-}\right\rangle  \notag \\
&=&\left\langle k_{+}\right\vert \hat{H}\left\vert m_{+}\right\rangle
\left\langle n_{-}\right\vert l_{-}\rangle -\left\langle k_{+}\right\vert
m_{+}\rangle \left\langle n_{-}\right\vert \hat{H}\left\vert
l_{-}\right\rangle  \notag \\
&&+i\gamma \left\langle k_{+}\right\vert \hat{a}\left\vert
m_{+}\right\rangle \left\langle n_{-}\right\vert \hat{a}^{\dag }\left\vert
l_{-}\right\rangle  \notag \\
&&-i\frac{\gamma }{2}\left\langle k_{+}\right\vert \hat{a}^{\dag }\hat{a}%
\left\vert m_{+}\right\rangle \left\langle n_{-}\right\vert l_{-}\rangle
\notag \\
&&-i\frac{\gamma }{2}\left\langle k_{+}\right\vert m_{+}\rangle \left\langle
n_{-}\right\vert \hat{a}^{\dag }\hat{a}\left\vert l_{-}\right\rangle .
\label{Hm}
\end{eqnarray}%
Equation (\ref{Hm}) shows that the operators act on the forward or backward
time branch, respectively. Therefore, we can introduce a generalized
Hamiltonian operator:%
\begin{equation}
\mathcal{\hat{H}}=\hat{H}_{+}-\hat{H}_{-}+i\gamma \hat{a}_{+}\hat{a}%
_{-}^{\dag }-i\frac{\gamma }{2}(\hat{a}_{+}^{\dag }\hat{a}_{+}+\hat{a}%
_{-}^{\dag }\hat{a}_{-}),  \label{H1}
\end{equation}%
where $\hat{H}_{\pm }$ are the Hamiltonians of the forward and backward time
branches, respectively. Based on Eqs.~(\ref{Hm}) and (\ref{H1}), $\mathcal{H}%
_{k,l,m,n}$ can be seen as a matrix element of $\mathcal{\hat{H}}$, i.e.,
\begin{equation}
\mathcal{H}_{k,l,m,n}=\left\langle n_{-}\right\vert \left\langle
k_{+}\right\vert \mathcal{\hat{H}}\left\vert l_{-}\right\rangle \left\vert
m_{+}\right\rangle ,
\end{equation}%
and the trace of $\hat{\rho}_{j+1}$ can thus be expressed as a simple form:
\begin{eqnarray}
&&\!\!\!\!\!\!\!\!\!\!\!\!\text{Tr}\hat{\rho}_{j+1}  \notag \\
&=&\!\!\!\text{Tr}\!\!\!\sum\limits_{k,l,m,n}\!\!\!\left( \delta
_{k,m}\delta _{l,n}-i\delta t\mathcal{H}_{k,l,m,n}\right) \left\langle
m_{+}\right\vert \hat{\rho}_{j}\left\vert n_{-}\right\rangle \left\vert
k_{+}\right\rangle \left\langle l_{-}\right\vert  \notag \\
&=&\!\!\!\text{Tr}\!\!\!\sum\limits_{k,l,m,n}\!\!\!\left( \delta
_{k,m}\delta _{l,n}-i\delta t\mathcal{H}_{k,l,m,n}\right) \left\vert
n_{-}\right\rangle \left\vert k_{+}\right\rangle \left\langle
l_{-}\right\vert \left\langle m_{+}\right\vert \hat{\rho}_{j}  \notag \\
&=&\!\!\!\text{Tr}\left( 1-i\delta t\mathcal{\hat{H}}\right) \hat{\rho}_{j}
\notag \\
&=&\!\!\!\text{Tr}\left( e^{-i\delta t\mathcal{\hat{H}}}\hat{\rho}%
_{j}\right) +O(\delta t^{2}).  \label{J}
\end{eqnarray}%
By iteration of Eq.~(\ref{J}), the density matrix can be evolved from $\hat{%
\rho}(t_{0})$ at $t_{0}$ to $\hat{\rho}(t_{f})$ at $t_{f}=t_{N}$. This
implies that in the limit $N\rightarrow \infty $ (and hence $\delta
t\rightarrow 0$),
\begin{equation}
Z_{t_{f},t_{0}}=\text{Tr}\hat{\rho}(t_{f})=\text{Tr}\left[ \exp (i\hat{S})%
\hat{\rho}(t_{0})\right] ,  \label{Z}
\end{equation}%
with $\hat{S}=-\int\nolimits_{t_{0}}^{t_{f}}\mathcal{\hat{H}}$d$t$.

Finally, we perform the limit, $t_{0}\rightarrow -\infty $ and $%
t_{f}\rightarrow +\infty $, to get the Keldysh partition function for
stationary states. Since in a Markov process, the initial state in the
infinite past does not affect the stationary state \cite{LM16}, we can
ignore the boundary term, i.e., $\hat{\rho}(t_{0})$ in Eq.~(\ref{Z}), and
obtain the final expression of the Keldysh partition function as%
\begin{equation}
Z=\text{Tr}\left[ \exp (i\hat{S})\right] ,
\end{equation}%
with the quantum action
\begin{equation}
\hat{S}=-\int\nolimits_{-\infty }^{+\infty }\mathcal{\hat{H}}\text{d}t.
\end{equation}

\section{Steady-state wave function for the nonlinear driving case}

We present the detailed derivation of Eq.~(\ref{WF2}) of the main text.
\begin{eqnarray}
\left\vert \Psi _{0}\right\rangle &=&e^{-\lambda \hat{a}_{cl}^{\dag
}}\left\vert \Phi _{0}\right\rangle =\frac{1}{\sqrt{N}}e^{-\lambda \hat{a}%
_{cl}^{\dag }}\left\vert 0\right\rangle _{q}\sum\limits_{k=0}^{+\infty }\phi
_{k}\left\vert k\right\rangle _{cl}  \notag \\
&=&\frac{1}{\sqrt{N}}\left\vert 0\right\rangle
_{q}\sum\limits_{j=0}^{+\infty }\frac{(-\lambda \hat{a}_{cl}^{\dag })^{j}}{j!%
}\sum\limits_{k=0}^{\infty }\phi _{k}\left\vert k\right\rangle _{cl}  \notag
\\
&=&\frac{1}{\sqrt{N}}\left\vert 0\right\rangle
_{q}\sum\limits_{j,k=0}^{+\infty }\frac{(-\lambda \hat{a}_{cl}^{\dag })^{j}}{%
j!}\frac{\left( 2\lambda \right) ^{k}}{\sqrt{k!}}\frac{(y)_{k}}{(z)_{k}}%
\left\vert k\right\rangle _{cl}  \notag \\
&=&\frac{1}{\sqrt{N}}\left\vert 0\right\rangle
_{q}\sum\limits_{j,k=0}^{+\infty }\frac{\left( 2\lambda \right)
^{k}(-\lambda )^{j}\sqrt{(j+k)!}}{j!k!}  \notag \\
&&\times \frac{(y)_{k}}{(z)_{k}}\left\vert j+k\right\rangle _{cl}.
\end{eqnarray}%
Let $j+k=m$, we obtain%
\begin{eqnarray}
\left\vert \Psi _{0}\right\rangle &=&\frac{1}{\sqrt{N}}\left\vert
0\right\rangle _{q}\sum\limits_{m=0}^{+\infty }\sum\limits_{k=0}^{m}\frac{%
\left( 2\lambda \right) ^{k}(-\lambda )^{m-k}m!}{\sqrt{m!}\left( m-j\right)
!k!}\frac{(y)_{k}}{(z)_{k}}\left\vert m\right\rangle _{cl}  \notag \\
&=&\frac{1}{\sqrt{N}}\left\vert 0\right\rangle
_{q}\sum\limits_{m=0}^{+\infty }\sum\limits_{k=0}^{m}\frac{\left( 2\lambda
\right) ^{k}(-\lambda )^{m-k}m!}{\sqrt{m!}\left( m-j\right) !k!}\frac{(y)_{k}%
}{(z)_{k}}\left\vert m\right\rangle _{cl}  \notag \\
&=&\frac{1}{\sqrt{N}}\left\vert 0\right\rangle
_{q}\sum\limits_{m=0}^{+\infty }\sum\limits_{k=0}^{m}\frac{2^{k}(-\lambda
)^{m}(-1)^{k}m!}{\sqrt{m!}\left( m-j\right) !k!}\frac{(y)_{k}}{(z)_{k}}%
\left\vert m\right\rangle _{cl}  \notag \\
&=&\frac{1}{\sqrt{N}}\left\vert 0\right\rangle
_{q}\sum\limits_{m=0}^{+\infty }\sum\limits_{k=0}^{+\infty }\frac{%
2^{k}(-\lambda )^{m}(-m)_{k}}{\sqrt{m!}k!}\frac{(y)_{k}}{(z)_{k}}\left\vert
m\right\rangle _{cl}  \notag \\
&=&\frac{1}{\sqrt{N}}\left\vert 0\right\rangle
_{q}\sum\limits_{m=0}^{+\infty }(-\lambda )^{m}\frac{_{2}F_{1}(-m,y;z;2)}{%
\sqrt{m!}}\left\vert m\right\rangle _{cl}.  \notag \\
&&
\end{eqnarray}

\end{document}